%% file: main.tex
\documentclass[9pt]{article}
\usepackage{spconf,amsmath,graphicx}
\usepackage{lipsum}
\usepackage{enumitem}
\setlist{nosep, leftmargin=14pt}

\usepackage{mwe} % to get dummy images
\usepackage{tabularx}
\usepackage{booktabs}
\usepackage{xcolor,colortbl}
\usepackage{multirow}
\usepackage{amsfonts}
\usepackage{bm}
\usepackage{graphicx}
\usepackage{subcaption}
\usepackage{overpic}
\usepackage{mathtools}
\usepackage{tabularray}
\usepackage{tikz}
\usepackage{rotating}
\usepackage{floatrow}
\usepackage{environ}
\usepackage{hyperref}
\newcolumntype{C}[1]{>{\centering\arraybackslash}p{#1}}
\definecolor{lavender}{rgb}{0.9, 0.9, 0.98}
\title{Direct Cardiac Segmentation from Undersampled K-space Using Transformers}
\name{Author(s) Name(s)}
\address{Author Affiliation(s)}
\name{
\begin{tabular}{@{}l@{}}
Yundi Zhang$^{\star{1,2}}$\thanks{$^\star$Authors contributed equally} \qquad 
Nil Stolt-Ansó$^{\star{1,3}}$ \qquad \qquad 
Jiazhen Pan$^{{1,2}}$\\ 
Wenqi Huang$^{{1,2}}$  \qquad 
Kerstin Hammernik $^{{1,4}}$ \qquad 
Daniel Rueckert$^{{1,2,3,4}}$
\end{tabular}
}

\address{
 $^{1}$ School of Computation, Information and Technology, Technical University Munich \\
$^{2}$ Klinikum Rechts der Isar, Technical University Munich \\ 
$^{3}$ Munich Center for Machine Learning, Technical University Munich\\
$^{4}$ Department of Computing, Imperial College London
}

\begin{document}
%\ninept
%
\maketitle
\begin{abstract}
The prevailing deep learning-based methods of predicting cardiac segmentation involve reconstructed magnetic resonance (MR) images. The heavy dependency of segmentation approaches on image quality significantly limits the acceleration rate in fast MR reconstruction. Moreover, the practice of treating reconstruction and segmentation as separate sequential processes leads to artifact generation and information loss in the intermediate stage. These issues pose a great risk to achieving high-quality outcomes. To leverage the redundant k-space information overlooked in this dual-step pipeline, we introduce a novel approach to directly deriving segmentations from sparse k-space samples using a transformer (DiSK). DiSK operates by globally extracting latent features from 2D+time k-space data with attention blocks and subsequently predicting the segmentation label of query points. We evaluate our model under various acceleration factors (ranging from 4 to 64) and compare against two image-based segmentation baselines. Our model consistently outperforms the baselines in Dice and Hausdorff distances across foreground classes for all presented sampling rates.

\end{abstract}

\input{chapters/1_introduction}

\input{chapters/2_method}

\input{chapters/3_experiments_and_result}

\input{chapters/5_conclusion}

\section{Acknowledgements}
This research study was conducted retrospectively using human subject data made available in open access by the UK Biobank Resource under Application Number 87802. Ethical approval was \textbf{not} required as confirmed by the license attached with the open access data.

This work is funded in part by the European Research Council (ERC) project Deep4MI (884622) and the Munich Center for Machine Learning. 

\bibliographystyle{IEEEbib}
\bibliography{strings,refs}

\end{document}

%% file: chapters/1_introduction.tex
\section{Introduction}
\label{sec:introduction}
Cardiac magnetic resonance (CMR) segmentation is vital for deriving key clinical metrics like ejection fraction and strain. The current deep learning-based workflow for the cardiac segmentation process is comprised of three key steps: data acquisition, reconstruction, and segmentation. In clinical practice, data acquisition involves sampling signals in frequency domain (k-space) to mitigate lengthy acquisition times caused by physiological and physical constraints. Reconstruction methods derive to restore the original MR images by exploiting the prior knowledge in either k-space~\cite{grappa, pan2023global} or image domain~\cite{otazo2015low, hammernik2018learning}. Research in cardiac segmentation primarily focuses on leveraging contextual inter-slice shape information~\cite{segshapeprior}, spatial details from adjacent slice~\cite{segspatial}, or temporal data~\cite{segtemporal}.

However, treating reconstruction and segmentation as separate serial tasks presents two major limitations. Firstly, errors and artifacts introduced during the earlier reconstruction phase can adversely impact the subsequent segmentation step, particularly prevalent at higher sampling rates. Secondly, the distinct objectives of reconstruction and segmentation may lead to a suboptimal solution for the final task. Conventionally, extensive care is taken to maximize the quality of images at pixel level during reconstruction, whereas segmentation tasks predominantly rely on high-level anatomical information presented in CMR images. Thus, viewing these two steps as a unified process holds great potential for enhancing segmentation performance at higher sampling rates.

Only a few works have endeavored to blur the boundary between reconstruction and segmentation. Schlemper et al.~\cite{schlemper2018cardiac} have trained segmentation models to be robust to reconstruction artifacts directly from zero-filled reconstructed images. Other networks~\cite{frnet, k2s} have fused reconstruction and segmentation processes through end-to-end training settings. Acar et al.~\cite{segawarerecon} have employed a segmentation model as a proxy task to emphasize anatomical structure during the reconstruction training. 

However, initiating the reconstruction from zero-filled k-space data introduces artifacts when employing the inverse fast Fourier transform (FFT). Furthermore, having an explicit image reconstruction step requires more information (in the form of k-space samples) than is required to directly infer a segmentation map. Notably, Rempe et al.~\cite{kstrip} have achieved skull segmentation directly from k-space data, bypassing an intermediate reconstruction by predicting masks in frequency domain. Nonetheless, they only use fully-sampled k-space to carry out the segmentation, which might deviate from standard clinical practices. 

\begin{figure*}[ht]
    \includegraphics[width=.8\textwidth]{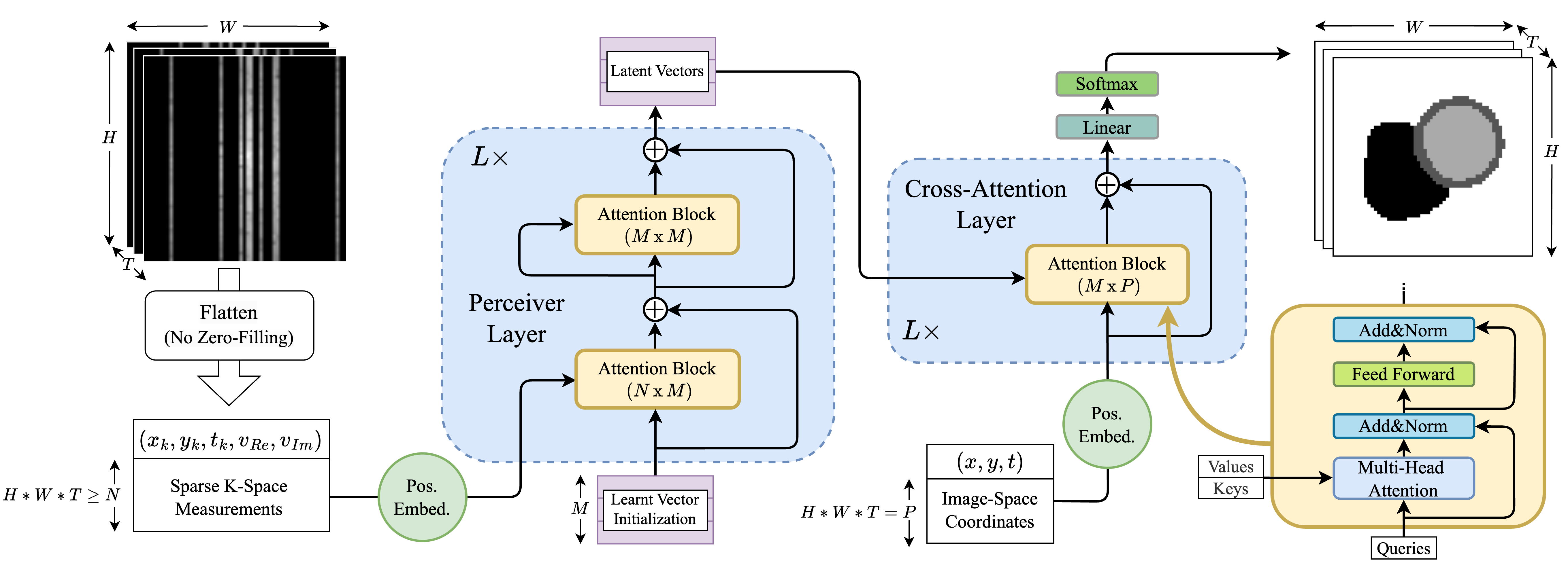}
    \centering
    \caption{Overview of DiSK architecture. $N$, $M$, and $P$ represent the number of sampled points in k-space, latent vectors, and query points in image domain, respectively. The encoder is made up of $L$ Perceiver layers~\cite{Jaegle2021} which alternates cross-attention and latent self-attention blocks. Specifically, cross-attention blocks project global features of the set of input k-space samples $K$ into a fixed-dimensional latent bottleneck of $M$ vectors. Self-attention between latent vectors contextualizes the extracted global features between the vectors. The decoder is made up of $L$ cross-attention layers which condition queried image-domain coordinates with the encoder's latent vectors into segmentation class probabilities.}
    \label{fig:architecture}
\end{figure*}

In this study, we propose DiSK, a novel approach to predict segmentation maps directly from undersampled (not zero-filled) k-space measurements using attention-based architecture. We demonstrate that operating from the source frequency data, bypassing the reconstruction step, is both sufficient and highly efficient in accurately recovering cardiac anatomy, even at high acceleration factors (up to 64). Our contributions can be summarized as follows:
\begin{itemize}
    \item We achieve 2D+time CMR image segmentation directly from undersampled k-space measurements \textbf{without} an intermediate image reconstruction step or an explicit inverse Fourier transform. 
    \item We utilize a transformer architecture to extract k-space features globally and subsequently predict segmentation maps in image space. 
    \item Our model shows superior Dice and Hausdorff distance scores compared to our baselines under acceleration factors ranging from 4 to 64.
\end{itemize}

%% file: chapters/2_method.tex
\section{Method}
\label{sec:method}
In this work, we utilize a transformer-based architecture that employs global attention to leverage all available k-space measurements into the output segmentation.

Due to the nature of the Fourier Transform (FT), local information in the image domain is correlated to the whole k-space domain. However, the convolutional kernels found in convolutional neural networks have an implicit bias towards local features. This puts a limit on their ability to integrate global information, causing them to inefficiently encode latent features from the k-space domain. 

On the other hand, we believe that transformers offer the necessary properties to appropriately process the nature of k-space: (1) the mechanism considers global correlations, (2) inputs of arbitrary sparsity caused by undersampling patterns are supported, and (3) feature extraction is invariant to the presented ordering of data samples. These properties offer the opportunity to process global relationships between k-space measurements, with no requirement to zero-fill or interpolate the input grid. 

% On the other hand, transformer architectures are based on attention mechanisms that place no assumption on the structure of the input set. Transformers treat inputs as sets of unordered tokens, removing any location-based restrictions on information processing. 

An overview of the architecture is presented in Figure~\ref{fig:architecture}. DiSK uses an encoder-decoder design to extract latent features from k-space and subsequently decodes coordinates into segmentation maps.  

We define the encoder as $E_{\bm{\theta}}(\left[V_k^u, X_k^u\right])$, where $V_k^u \in \mathbb{C}^{N\times1}$ is the set of $N$ sampled k-space values, $X_k^u \in \mathbb{R}^{N\times3}$ are the coordinates of each given k-space value, and $[\cdot, \cdot]$ represents element-wise concatenation. We use the $u$ superscript to denote the undersampled nature of the input set $\left|V_k^u\right| \leq \left|V_k\right|$. The encoder uses the Perceiver architecture~\cite{Jaegle2021} to iteratively aggregate global k-space information. The alternating design between cross-attention (CA) and self-attention (SA) blocks circumvents the $\mathcal{O}(N^2)$ memory complexity of naive SA between input tokens of standard transformers. Given that the number of k-space samples in our data ranges from tens to hundreds of thousands of points, this architecture allows us to efficiently scale up the number of sample points that can be processed. The CA blocks project global k-space information into a fixed set of latent vectors $H \in \mathbb{R}^{M\times d}$, while the self-attention (SA) blocks contextualize features between latent vectors.

The decoder $D_{\bm{\phi}}(\bm{h}, \bm{x})$ is comprised of a sequence of CA blocks that allow for any queried image-domain coordinate $\bm{x} \in \mathbb{R}^3$ to be conditioned by the latent vectors into decoding a probabilistic segmentation vector $\hat{\bm{s}}_{\bm{x}} \in \mathbb{R}^C$ of $C$ classes. The segmentation output is supervised using Dice and binary cross-entropy (BCE) losses. The overall model $G$ can be formulated as:
\begin{equation}
    G_{\bm{\Theta}}(\left[X_k^u, V_k^u\right], \bm{x}) = D_{\bm{\phi}}(E_{\bm{\theta}}(\left[X_k^u, V_k^u\right]), \bm{x}) = \hat{\bm{s}}_{\bm{x}}\:,
\end{equation}
where $\hat{\bm{s}}_{\bm{x}}$ is the predicted segmentation class probabilities.

We apply the positional encoding outlined in~\cite{mildenhall2021nerf} individually to each coordinate and value element, which increases the network's sensitivity to small changes in the input.

%% file: chapters/3_experiments_and_result.tex
\section{Experiments and result}
\label{sec:experiments_and_result}

\textit{\textbf{Dataset.}} Our dataset is comprised of 1200 mid-ventricular slices of cardiac short-axis scans from the UK-Biobank study~\cite{petersen2015uk}. The dataset is arranged into 1000/100/100 splits for training, validation, and testing sets. Each scan has 50 frames with an average image size of approximately 80x80 pixels per frame. As only the magnitude images were available, we create synthetic k-space data for each 2D+time scan by applying additional Gaussian B0 variations on the fly in order to remove the conjugate symmetry of k-space. Similarly, Cartesian undersampling masks are generated in real-time independently for each frame by sampling normally distributed lines centered on the DC component.

\textit{\textbf{Implementation details.}}
We set the number of layers $L$ in the transformer to 4 and token sizes to 128 for both the encoder and decoder. Fully-connected layers at each attention block had 128 nodes. We use a total of 128 latent vectors (also of size 128), where initialization prior to the first layer is optimizable with respect to the overall loss. The segmentation head is a linear layer followed by a softmax non-linearity for 4 classes. We conducted all experiments on a single NVIDIA RTX A4000 GPU using the Pytorch library. Code for training our model is made
publicly available\footnote{https://github.com/Yundi-Zhang/DiSK.git}.

\textit{\textbf{Baselines.}}
To the best of our knowledge, no previous work has attempted to produce cardiac segmentations directly from k-space samples. We thus implement two baselines from ~\cite{schlemper2018cardiac} that have shown promising cardiac segmentation performance on images reconstructed from high acceleration factors via zero-filling: Syn-Net and LI-Net.

\begin{figure}[h]
    \resizebox{\columnwidth}{!}{%
    % \begin{tikzpicture}[scale=.99\columnwidth/20cm, font=\small] 
    \begin{tikzpicture}[font=\LARGE]
    % \draw[step=1cm,gray,very thin] (0,0) grid (20, 17);
    %label
    \node[anchor=north](Acc.) at (-0.3,14.4) {Acc.};
    \node[anchor=north](GT) at (1.6,14.4) {GT};
    \node[anchor=north](Syn-Net) at (4.7,14.4) {Syn-Net};
    \node[anchor=north](Li-Net) at (7.8,14.4) {LI-Net};
    \node[anchor=north](Ours) at (10.7,14.4) {Ours};
    % \node[anchor=north](Recon) at (13.8,17.1) [align=center]{Zero-filled\\ Recon.};
    \node[anchor=north](k-space) at (13.6,14.4) {Zero-filling};
    % 1st row
    \node[anchor=east](4x) at (0.2,12.26) {$4 \times$};
    \node[anchor=south west, scale=1.06] (4x/gt) at (0,10.96)
    {\includegraphics[width=2.8cm]{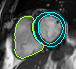}};
    \node[anchor=south west, scale=1.06] (4x/unet) at (3,10.96)
    {\includegraphics[width=2.8cm]{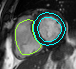}};
    \node[anchor=south west, scale=1.06] (4x/linet) at (6,10.96)
    {\includegraphics[width=2.8cm]{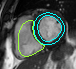}};
    \node[anchor=south west, scale=1.06] (4x/ours) at (9,10.96)
    {\includegraphics[width=2.8cm]{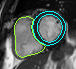}};
    % \node[anchor=south west, scale=1.06] (4x/recons) at (12,10.96)
    % {\includegraphics[width=2.8cm]{Figures/acc/recon_4.png}};
    \node[anchor=south west, scale=1.06] (4x/kspace) at (12,10.96)
    {\includegraphics[width=2.8cm]{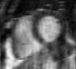}};
    % 2nd row
    \node[anchor=east](8x) at (0.2,9.52) {$8 \times$};
    \node[anchor=south west, scale=1.06] (8x/gt) at (0,8.22)
    {\includegraphics[width=2.8cm]{Figures/acc/gt.png}};
    \node[anchor=south west, scale=1.06] (8x/unet) at (3,8.22)
    {\includegraphics[width=2.8cm]{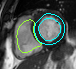}};
    \node[anchor=south west, scale=1.06] (8x/linet) at (6,8.22)
    {\includegraphics[width=2.8cm]{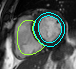}};
    \node[anchor=south west, scale=1.06] (8x/ours) at (9,8.22)
    {\includegraphics[width=2.8cm]{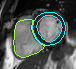}};
    % \node[anchor=south west, scale=1.06] (8x/recons) at (12,8.22)
    % {\includegraphics[width=2.8cm]{Figures/acc/recon_8.png}};
    \node[anchor=south west, scale=1.06] (8x/kspace) at (12,8.22)
    {\includegraphics[width=2.8cm]{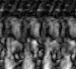}};
    % 3rd row
    \node[anchor=east](16x) at (0.2,6.78) {$16 \times$};
    \node[anchor=south west, scale=1.06] (16x/gt) at (0,5.48)
    {\includegraphics[width=2.8cm]{Figures/acc/gt.png}};
    \node[anchor=south west, scale=1.06] (16x/unet) at (3,5.48)
    {\includegraphics[width=2.8cm]{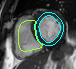}};
    \node[anchor=south west, scale=1.06] (16x/linet) at (6,5.48)
    {\includegraphics[width=2.8cm]{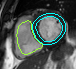}};
    \node[anchor=south west, scale=1.06] (16x/ours) at (9,5.48)
    {\includegraphics[width=2.8cm]{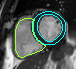}};
    % \node[anchor=south west, scale=1.06] (16x/recons) at (12,5.48)
    % {\includegraphics[width=2.8cm]{Figures/acc/recon_16.png}};
    \node[anchor=south west, scale=1.06] (16x/kspace) at (12,5.48)
    {\includegraphics[width=2.8cm]{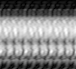}};
    % 4th row
    \node[anchor=east](32x) at (0.2,4.04) {$32 \times$};
    \node[anchor=south west, scale=1.06] (32x/gt) at (0, 2.74)
    {\includegraphics[width=2.8cm]{Figures/acc/gt.png}};
    \node[anchor=south west, scale=1.06] (32x/unet) at (3,2.74)
    {\includegraphics[width=2.8cm]{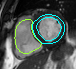}};
    \node[anchor=south west, scale=1.06] (32x/linet) at (6,2.74)
    {\includegraphics[width=2.8cm]{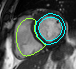}};
    \node[anchor=south west, scale=1.06] (32x/ours) at (9,2.74)
    {\includegraphics[width=2.8cm]{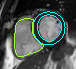}};
    % \node[anchor=south west, scale=1.06] (32x/recons) at (12,2.74)
    % {\includegraphics[width=2.8cm]{Figures/acc/recon_32.png}};
    \node[anchor=south west, scale=1.06] (32x/kspace) at (12,2.74)
    {\includegraphics[width=2.8cm]{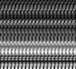}};
    
    % 5th row
    \node[anchor=east](64x) at (0.2,1.3) {$64 \times$};
    \node[anchor=south west, scale=1.06] (64x/gt) at (0,0)
    {\includegraphics[width=2.8cm]{Figures/acc/gt.png}};
    \node[anchor=south west, scale=1.06] (64x/unet) at (3,0)
    {\includegraphics[width=2.8cm]{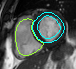}};
    \node[anchor=south west, scale=1.06] (64x/linet) at (6,0)
    {\includegraphics[width=2.8cm]{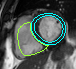}};
    \node[anchor=south west, scale=1.06] (64x/ours) at (9,0)
    {\includegraphics[width=2.8cm]{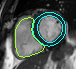}};
    % \node[anchor=south west, scale=1.06] (64x/recons) at (12,0)
    % {\includegraphics[width=2.8cm]{Figures/acc/recon_64.png}};
    \node[anchor=south west, scale=1.06] (64x/kspace) at (12,0)
    {\includegraphics[width=2.8cm]{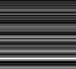}};
    
    \end{tikzpicture}
    }
    \caption{Test set segmentations predicted from different models over varying acceleration factors. The last column visualizes the undersampled k-space measurements in a time frame.}
    \label{fig:acc}
    \end{figure}

\textit{\textbf{Results.}}
We train and evaluate all models under acceleration factors (4, 8, 16, 32, 64). Performances are quantified by taking average Dice over the three foreground classes, as well as the maximum Hausdorff distance over the foreground classes. Results are reported in Table~\ref{table:scores}, where the scores are averaged over three foreground classes to enhance better anatomical accuracy. Figure~\ref{fig:acc} shows a qualitative example of all models predicting segmentations from sparse k-space measurements. The corresponding undersampled k-space measurements and zero-filled reconstructed images are shown on the side. Our model consistently outperforms the baselines by approximately 0.1 in Dice scores and demonstrates a reduction of roughly 2 pixels in Hausdorff distances across foreground classes for all presented sampling rates.
 
    \begin{table}[ht]
    \centering
    \caption{Dice scores and Hausdorff distances over a testing set of 100 subjects.}
    \resizebox{\columnwidth}{!}{
    \begin{tabular}{c c c c c}
    \toprule \multirow{2}{*}{  } Acc. &   & Syn-Net & LI-Net &  Ours \\
    \hline \rowcolor[gray]{0.9}  & Dice $\uparrow$ & $0.749_{ \pm 0.260}$ & $0.805_{ \pm 0.198}$ & $\bm{0.902_{ \pm 0.089}}$ \\
    \rowcolor[gray]{0.9} \multirow{-2}{*}{4$\times$} & HD \: $\downarrow$& $6.809_{ \pm 2.776}$ & $6.557_{ \pm 3.160}$ & $\bm{4.797_{ \pm 2.064}}$ \\
    \hline & Dice $\uparrow$ & $0.748_{ \pm 0.258}$ & $0.809_{ \pm 0.192}$ & $\bm{0.902_{ \pm 0.089}}$ \\
    \multirow{-2}{*}{8$\times$} & HD \: $\downarrow$ & $6.794_{ \pm 2.868}$ & $7.019_{ \pm 3.521}$ & $\bm{4.772_{ \pm 2.253}}$ \\
    \hline \rowcolor[gray]{0.9}  & Dice $\uparrow$ & $0.742_{ \pm 0.264}$ & $0.800_{ \pm 0.197}$ & $\bm{0.903_{ \pm 0.085}}$ \\
    \rowcolor[gray]{0.9} \multirow{-2}{*}{16$\times$} & HD \: $\downarrow$ & $6.792_{ \pm 2.818}$ & $6.841_{ \pm 2.971}$ & $\bm{4.509_{ \pm 2.068}}$ \\
    \hline & Dice $\uparrow$ & $ 0.723_{ \pm 0.287}$ & $0.752_{ \pm 0.242}$ & $\bm{0.902_{ \pm 0.086}}$ \\
    \multirow{-2}{*}{32$\times$} & HD \: $\downarrow$ & $7.383_{ \pm 3.122}$ & $7.531_{ \pm 3.131}$ & $\bm{4.665_{ \pm 2.054}}$ \\
    \hline \rowcolor[gray]{0.9}  & Dice $\uparrow$ & $0.733_{ \pm 0.261}$ & $0.799_{ \pm 0.190}$ & $\bm{0.902_{ \pm 0.085}}$ \\
    \rowcolor[gray]{0.9} \multirow{-2}{*}{64$\times$} & HD \: $\downarrow$ & $7.543_{ \pm 2.972}$ & $6.706_{ \pm 2.567}$ & $\bm{4.911_{ \pm 2.356}}$ \\
    \bottomrule
    \end{tabular}
    }
    \label{table:scores}
\end{table}    
    % \begin{figure}
    
    % \begin{subfigure}{.5\textwidth}
    %   \centering
    %   \includegraphics[width=\columnwidth]{Figures/dice.png}
    % \end{subfigure}%
    % \begin{subfigure}{.5\textwidth}
    %   \centering
    %   \includegraphics[width=\columnwidth]{Figures/hd.png}
    % \end{subfigure}
    %     \caption{Distribution of Dice scores and Hausdorff distances over testing set.}
    % \label{fig:metrics}
    % \end{figure}

\textit{\textbf{Discussion.}}
In summary, we propose a network aimed at deriving segmentation from undersampled k-space data, omitting the image reconstruction stage. We demonstrate the superior performance of DiSK compared to our baselines on 2D+time cardiac dataset. 

Interestingly, both our approach and the baselines show little degradation under the higher acceleration rates. We believe the models are prone to memorizing the general segmentation labels, by either leveraging the image borders in the case of the image-based baselines, or by correlating coordinates with segmentation labels like in the case of our model. In some sense, the models can be seen as memorizing an implicit prior over the typical anatomy in the dataset and learning how to deform its default prediction with subject-specific information.

In future work, we aim to expand upon the current work to conduct full-view short-axis cardiac MRI by processing larger numbers of k-space samples. We also hope to investigate how the reliance on shape memorization is influenced by the introduction of rotations and translations of the input. Moreover, we would like to explore other downstream tasks such as cardiac function assessment and disease classification.

%% file: chapters/5_conclusion.tex
\section{Conclusion}
\label{sec:conclusion}

To the best of our knowledge, this is the first study that explores the direct prediction of segmentation maps directly from sparse under-sampled k-space measurements. Compared to conventional segmentation models, the proposed model shows a novel data processing pipeline for medical downstream tasks, where global feature extraction is learnt directly from k-space data without the potential introduction of artifacts or the loss of information due to intermediate image reconstruction steps.